## *Vavilov-Cerenkov phenomenon in metal nanofilms*

V. S. Zuev

P.N.Lebedev Physical Institute of RAS

53, Leninsky pr., Moscow 119991

The Vavilov-Cerenkov phenomenon – light emission during the motion of a fast electron in a dense medium is observable in homogeneous media as well as in inhomogeneous media like photonic crystals. So long as a space (homogeneous or inhomogeneous) contains natural waves with phase velocities that are lower than the velocity of a flying through electron then a Cerenkov emission of light would appear.

In nanofilms and nanowires made of Ag, Au, Cu there are so called surface plasmon polaritons, namely natural waves with a low phase velocity. These waves velocity could be by tens and hundreds times lower than the light velocity in vacuum. An electron with a velocity that is comparatively lower than a velocity of an electron that is able to emit in a uniform medium could emit Cerenkov light. Dispersion relations are calculated for the corresponding plasmons as well as the angles of emission of the plasmons of corresponding frequencies.

## *Явление Вавилова-Черенкова в нанопленке металла*

В.С.Зуев

Физический институт им. П.Н.Лебедева РАН

119991 Москва, Ленинский пр-т, 53

vizuev@sci.lebedev.ru

Явление Вавилова-Черенкова – излучение света при движении быстрого электрона в среде наблюдают как в однородных средах, так и в неоднородных, в таких, как фотонные кристаллы. Условием возникновения этого излучения является наличие в пространстве (однородном или неоднородном) собственных электромагнитных волн с фазовой скоростью меньше, чем скорость пролетающего электрона.

В нанопленках и в нанонитях из серебра, золота, меди имеются так называемые поверхностные плазмон-поляритоны, собственные волны с малой фазовой скоростью. Отличие фазовой скорости этих волн от скорости света в вакууме может достигать многих десятков и сотен раз. Испускать излучение в виде плазмона будет электрон, сравнительно медленный по сравнению с электроном, способным излучать в однородной среде. Рассчитаны дисперсионные соотношения для соответствующих плазмонов и углы испускания плазмонов соответствующих частот.

### *Явление Вавилова-Черенкова в нанопленке металла*

В.С.Зуев

Физический институт им. П.Н.Лебедева РАН

119991 Москва, Ленинский пр-т, 53

vizuev@sci.lebedev.ru

Явление Вавилова-Черенкова – излучение света при движении быстрого электрона в среде наблюдают как в однородных средах /1,2/, так и в неоднородных, в таких, как фотонные кристаллы /3/. Условием возникновения этого излучения является наличие в пространстве (однородном или неоднородном) собственных электромагнитных волн с фазовой скоростью меньше, чем скорость пролетающего электрона.

Собственные волны с малой фазовой скоростью имеются в нанопленках и в нанонитях из серебра, золота, меди. Это так называемые поверхностные плазмон-поляритоны, или для краткости – поверхностные плазмоны. Отличие фазовой скорости этих волн от скорости света в вакууме может достигать многих десятков и сотен раз. Это означает, что испускать излучение в виде плазмона будет электрон, сравнительно медленный по сравнению с электроном, способным излучать в однородной среде.

В пространстве с тонкой металлической пленкой имеются симметричный и антисимметричный поверхностные плазмона. Классификацию плазмонов мы делаем по виду магнитного поля. При заданном значении волнового числа частота симметричного плазмона выше частоты антисимметричного плазмона. Поле плазмонов локализовано на пленке, а на удалении от пленки экспоненциально мало. Таких волн в однородном пространстве нет. Дисперсионные соотношения для плазмонов на пленке имеют вид /4,5/

$$k^2 = \left( \frac{\sqrt{k^2 - (\omega/c)^2 \varepsilon_2}}{\varepsilon_2} \frac{1 \mp \exp[-d\sqrt{k^2 - (\omega/c)^2 \varepsilon_2}]}{1 \pm \exp[-d\sqrt{k^2 - (\omega/c)^2 \varepsilon_2}]} \right)^2 + (\omega/c)^2. \quad (1)$$

Верхние знаки относятся к симметричному плазмону, нижние – к антисимметричному плазмону. Диэлектрическую проницаемость металла $\varepsilon_2$ следует считать равной

$$\varepsilon_2(\omega) = 1 - \frac{\omega_{pl}^2}{\omega^2}, \quad \omega_{pl} = \sqrt{\frac{4\pi n e^2}{m_e}} = 5.64 \cdot 10^4 \sqrt{n_e} \ \text{рад/с}. \quad (2)$$

Квадрат волнового числа $k^2$ представляет собой сумму $k_x^2 + k_y^2$. Оси $x$ и $y$ лежат в плоскости пленки, ось $z$ нормальна к пленке. Плазмоны представляют собой поперечно-магнитные волны. Они не имеют продольной магнитной составляющей.

Типичные дисперсионные кривые приведены на рис.1. Они рассчитаны для пленки толщиной 10 нм при $\omega_{pl} \approx 1.346 \cdot 10^{16} \ c^{-1}$ ($n = 5.7 \cdot 10^{22} \ cm^{-3}$).

Среди многочисленных поверхностных мод металлического цилиндра только так называемая $TM_0$ мода не имеет критической частоты: она существует на цилиндрах сколь угодно малого диаметра.

Поле этой моды имеет вид ($r, \theta, z$ - цилиндрические координаты) /6/

$$\vec{N}_{0\gamma} = \vec{n}_{0\gamma} e^{ihz}, \quad \vec{n}_{0\gamma} = i \frac{h}{\sqrt{k^2}} \frac{d}{dr} Z_0(\gamma r) \cdot \vec{i}_r - \frac{\gamma^2}{\sqrt{k^2}} Z_0(\gamma r) \cdot \vec{i}_z. \quad (3)$$

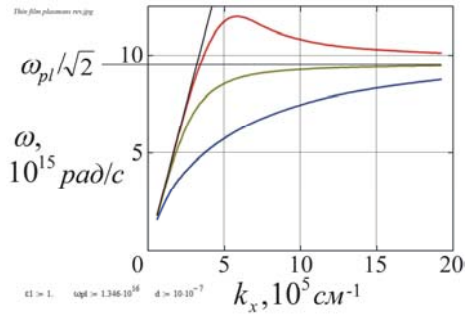
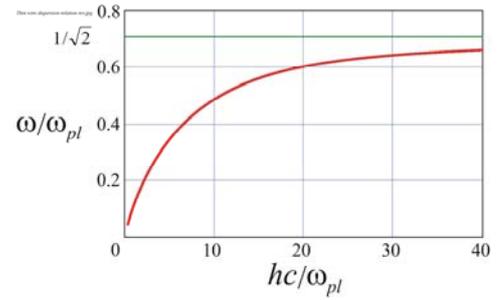

Рис.1. Дисперсионные кривые для поверхностных плазмонов на тонкой пленке металла в вакууме. Красная кривая – симметричный плазмон, синяя кривая – антисимметричный плазмон, черная горизонтальная прямая - $\omega_{pl}/\sqrt{2}$, наклонная черная прямая – световая линия, $\omega = kc$. Толщина пленки 10 нм, $\omega_{pl} \approx 1.346 \cdot 10^{16}\, c^{-1}$.

Рис.2. Дисперсионная кривая для поверхностного плазмона $TM_0$ на металлической нити.

Здесь $\gamma = \sqrt{h^2 - k^2}$, $k^2 = \varepsilon\mu(\omega/c)^2$, $Z_0(\gamma r)$ равно либо $I_0(\gamma r)$, либо $K_0(\gamma r)$, обе – модифицированные бесселевы функции. Поскольку $k^2$ в данной задаче принимает как положительное, так и отрицательное значение, то $\sqrt{k^2}$ равен либо $k$, если $k^2 > 0$, либо $i|k|$, если $k^2 < 0$. В соответствии со свойствами модифицированных бесселевых функций для пространства внутри цилиндра выбираем $I_0(\gamma r)$, которая ограничена при $r = 0$, для пространства вне цилиндра - $K_0(\gamma r)$, которая убывает монотонно с ростом $r$. Составив граничные условия равенства тангенциальных компонент полей на границе $r = a$, получаем характеристическое уравнение ($\mu_1 = \mu_2 = 1$) /4/

$$\frac{\gamma_2}{\gamma_1}\frac{I_0(\gamma_2 a)}{K_0(\gamma_1 a)} = -\frac{\varepsilon_2}{\varepsilon_1}\frac{I_1(\gamma_2 a)}{K_1(\gamma_1 a)}. \tag{4}$$

В (4) будем считать $\varepsilon_1 = 1$ и $\varepsilon_2 = 1 - \omega_{pl}^2/\omega^2$. На рис.2 дисперсионное соотношение (4) представлено в виде графика.

Черенковское излучение возникает в том случае, когда в пространстве имеются собственные электромагнитные волны с фазовой скоростью, которая меньше скорости электрона. В однородном пространстве, заполненном диэлектриком с $\varepsilon > 1$, это однородные плоские волны $\vec{A}e^{i\vec{k}\cdot\vec{r}-i\omega t}$, в фотонном кристалле – соответствующие фотонные волны, собственные моды кристалла /3/, в нанопленках и нанонитях металла – поверхностные плазмон-поляритоны без критических частот, рассмотренные выше.

В классическом случае в эффекте Вавилова-Черенкова в однородном пространстве имеет место следующее соотношение /7/:

$$\frac{c}{v} = \sqrt{\varepsilon}\cos\theta \rightarrow \frac{(\omega/v)}{\cos\theta} = k = \sqrt{\varepsilon}(\omega/c), \tag{5}$$

где $v$ - скорость частицы. Применительно к неоднородным пространствам правая часть последнего равенства должна быть заменена на $k$, которое для избранной частоты $\omega$ следует из дисперсионного соотношения $k = \varphi(\omega)$ для собственных волн рассматриваемого неоднородного пространства /3/. В рассматриваемых случаях это (1) для плазмонов на пленке и (4) для плазмонов на нити. Численные расчеты дают результаты, изображенные на рис.3 и 4 (3 – для пленки, 4 – для нити).

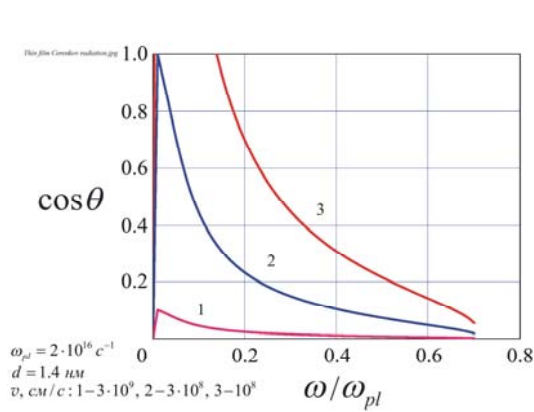
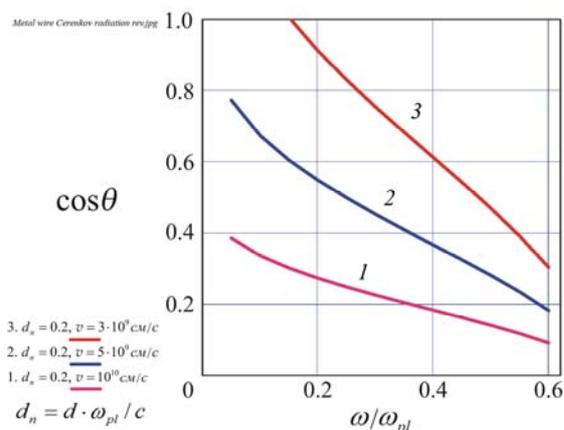

Рис.3. Излучение Черенкова в металлической пленке толщины $1.4$ нм, $\omega_{pl} = 2 \cdot 10^{16}\ c^{-1}$. Параметр – скорость электрона: 1 - $3 \cdot 10^9$ см/с, 2 - $3 \cdot 10^8$ см/с, 3 - $10^8$ см/с.

Рис.4. Излучение Черенкова в металлической нити диаметра $3$ нм, $\omega_{pl} = 2 \cdot 10^{16}\ c^{-1}$. Параметр – скорость электрона: 1. $10^{10}$ см/с, 2. $5 \cdot 10^9$ см/с, 3. $3 \cdot 10^9$ см/с.

Электрон может пронизывать пленку под любым углом к ее поверхности. Электрон должен пересекать нить под углом к ее оси. Электрон, движущийся по оси нити не будет возбуждать плазмон в нити, так как ни для одной из волн нити не возникнет пространственного синхронизма.

Собственные волны наносферы, наносфероида и других наночастиц являются стоячими волнами. Стоячая волна как минимум состоит из двух бегущих волн. Движущийся электрон, находящийся в пределах собственной волны наночастицы, будет возбуждать попутную волну. Встречная волна не будет возбуждаться непосредственно электроном, а будет возникать благодаря отражению попутной волны на границе наночастицы. Волна в частице будет возбуждаться в том случае, когда соответствующие $k$ и $\omega$ будут соответствовать волнам, присутствующим в спектре волн движущегося электрона. Для наночастицы несферической формы еще необходим подходящий угол пролета электрона.

Вернемся к плазмонам на нанопленке металла. Скорость $3 \cdot 10^8$ см/с меньше фермиевской скорости в серебре, где энергия Ферми /8/ равна $E_f = 5.5$ эВ (по расчету) /9/. Это означает, что электроны проводимости в серебре с энергией вблизи $E_f$ будут порождать черенковское излучение в виде поверхностных плазмонов. На это обстоятельство обратил внимание А.П.Канавин (6 февраля 2009 г.). Испускание поверхностных плазмонов электронами проводимости в тонкой пленке металла и последующее их поглощение металлом оказывается дополнительным механизмом релаксации энергии в металле.

Устройство с пленкой, с нитью, с наночастицей, все из металла, может быть основой детектора движущихся электронов. В отличие от черенковского детектора классического устройства с однородным диэлектриком это новое устройство способно детектировать менее энергичные электроны, тогда как классическое устройство детектирует электроны с энергией в сотни кэВ.

Поверхностные плазмоны не могут быть зарегистрированы внешним фотоприемником. Однако предложены средства трансформации плазмонов в электромагнитные волны свободного пространства /10/. Впрочем, любые неоднородности на поверхности тела с плазмонами будут порождать распространяющиеся волны в окружающем пространстве, и вопрос заключается лишь в эффективности такого способа преобразования.

Подведем итог. Движущийся электрон, пересекающий в своем движении либо нанопленку, либо нанонить, либо наночастицу, все – из металла, чья диэлектрическая проницаемость отрицательна, будет возбуждать поверхностные плазмон-поляритоны. Условием возбуждения является выполнение условия $v_{эл} > v_{фаз}$, где $v_{эл}$ - скорость электрона, а $v_{фаз}$ - скорость поверхностной волны. Последняя может отличаться в десятки и сотни раз от фазовой скорости электромагнитной волны в свободном пространстве. В результате возникает возможность детектирования электронов со сравнительно малой скоростью, соответствующей энергии электрона в несколько $эВ$, в отличие от энергии электронов в сотни $кэВ$, детектируемых в традиционных устройствах с возбуждением черенковского излучения. Малоэнергичные электроны могут детектироваться по черенковскому излучению в устройствах с фотонными кристаллами. Устройства с поверхностными плазмонами являются альтернативой устройствам с фотонными кристаллами.

1. П.А.Черенков. Труды ФИАН, т. 2, вып 4, стр 3-62 (1944)
2. Л.Д.Ландау, Е.М.Лифшиц. Электродинамика сплошных сред. Москва, Наука, 1982
3. C.Luo, M.Ibanescu, S.G.Johnson, J.D.Joannopoulos. Science, v.299, pp.368-371 (2003)
4. V.S.Zuev, G.Ya.Zueva. J. Russian Laser Res., v.27, pp.167-184 (2006)
5. V.S.Zuev, G.Ya.Zueva. arXiv:0705.2529v3 24 May 2007
6. J.A.Stratton. Electromagnetic Theory. McGraw-Hill Book Co., N.Y., London, 1941
7. И.Е.Тамм, И.М.Франк. ДАН СССР, т.14, 107-112 (1937)
8. Ch. Kittel, C.Y.Fong. Quantum Theory of Solids, 1963
9. *www.lgrflab.ru/physbook/tom5/ch6/texthtml/ch6_5_text.htm*, расчет для $n_e = 5 \cdot 10^{22} \, см^{-3}$
10. X.Y.Yang, H.T.Liu, P.Lalanne. Phys. Rev. Lett., v.102, 153903 (2009)